\documentclass[letterpaper, 10 pt, conference]{ieeeconf}  

\IEEEoverridecommandlockouts                              

\usepackage{amsmath,amssymb,amsfonts}
\usepackage{algorithmic}
\usepackage{graphicx}
\usepackage{subcaption}
\usepackage{algorithm,algorithmic}
\usepackage{hyperref}
\usepackage{newtxtext,newtxmath}
\usepackage{multirow}
\usepackage{textcomp}
\usepackage{booktabs}
\usepackage{colortbl}
\title{\LARGE \bf Multimodal Sleep Stage and Sleep Apnea Classification Using Vision
Transformer: A Multitask Explainable Learning Approach
 }

\author{Kianoosh Kazemi$^{1}$, Iman Azimi$^{2}$, Michelle Khine$^{3}$, Rami N. Khayat$^{4}$, Amir M. Rahmani$^{2}$ and Pasi Liljeberg$^{1}$
\thanks{*This work was not supported by any organization}
\thanks{$^{1}$Kianoosh Kazemi and Pasi Liljeberg are with the Department of 
Computing, University of Turku,  Finland
        {\tt\small kianoosh.k.kazemi@utu.fi and pasi.liljeberg@utu.fi}}%
\thanks{$^{2}$Amir M. Rahmani and Iman Azimi are  with the Department of Computer Science, University of California, Irvine, Irvine, CA, United States
        {\tt\small a.rahmani@uci.edu and azimii@uci.edu}}%
\thanks{$^{3}$Michelle Khine is with the Department of Biomedical Engineering, University of California Irvine, Irvine, CA, United States
        {\tt\small mkhine@uci.edu}}%
\thanks{$^{4}$Rami N. Khayat is with the Division of Pulmonary and Critical Care Medicine, The UCI Comprehensive Sleep Center, University of California. Irvine, Newport Beach, CA, United States
        {\tt\small khayatr@hs.uci.edu}}%
}

\begin{document}

\maketitle
\thispagestyle{empty}
\pagestyle{empty}

\begin{abstract}
Sleep is an essential component of human physiology, contributing significantly to overall health and quality of life. Accurate sleep staging and disorder detection are crucial for assessing sleep quality. Studies in the literature have proposed PSG-based approaches and machine-learning methods utilizing single-modality signals. However, existing methods often lack multimodal, multilabel frameworks and address sleep stages and disorders classification separately. In this paper, we propose a 1D-Vision Transformer for simultaneous classification of sleep stages and sleep disorders. Our method exploits the sleep disorders' correlation with specific sleep stage patterns and performs a simultaneous identification of a sleep stage and sleep disorder. The model is trained and tested using multimodal-multilabel sensory data (including photoplethysmogram, respiratory flow, and respiratory effort signals). The proposed method shows an overall accuracy (cohen's Kappa) of 78\% (0.66) for five-stage sleep classification and 74\% (0.58) for sleep apnea classification. Moreover, we analyzed the encoder attention weights to clarify our models’ predictions and investigate the influence different features have on the models’ outputs. The result shows that identified patterns, such as respiratory troughs and peaks, make a higher contribution to the final classification process.
\end{abstract}

\section{INTRODUCTION}
Sleep is an integral part of human life and plays a crucial role in the well-being of the body \cite{c1}. Sleep plays a crucial role in the recovery of physical and cognitive functions \cite{c2}. There are several stages in sleep: non-rapid eye movement (NREM) and rapid eye movement (REM) \cite{c4}, each marked by distinct brain activity patterns and physiological changes \cite{c3}.  In NREM sleep, three stages are distinguished (N1, N2, and N3) according to specific patterns of brain waves. REM sleep is characterized by rapid eye movements, vivid dreaming, and physiological changes such as rapid heart rate and breathing \cite{c5}. Thus, accurate sleep stages detection and analysis are crucial parts of general health assessment.  

Polysomnography (PSG) is the standard clinical approach to sleep staging sleep staging \cite{c6}. This technique provides detailed insights into sleep patterns and disorders by collecting multimodal signals, including electroencephalograms (EEG), electrocardiograms (ECG), electrooculography (EOG), and electromyograms (EMG) data \cite{c6}. The process involves patients sleeping in a lab environment while two clinicians manually analyze the recorded data to identify sleep stages and potential anomalies. Despite its precision, PSG is labor-intensive, time-consuming, and limited to controlled settings, making it impractical for longitudinal monitoring \cite{c18}. In recent years, wearable devices such as the Oura Ring and Fitbit have emerged as convenient and non-invasive alternatives for home-based sleep tracking. These wearables utilize photoplethysmography (PPG) alongside other sensors, including those for temperature and respiration, to estimate sleep duration and stages \cite{c20, c21}. Although they offer cost-effective and scalable solutions, their accuracy in sleep staging is inferior to PSG, with clinical validation studies questioning their reliability \cite{c22}. 


Wearable sensing technology and deep learning (DL) have advanced sleep staging research, particularly home monitoring. Several studies leveraged deep learning approaches for sleep staging using wearable data, such as PPG, heart rate variability, and respiratory signals \cite{c23, c26}. 
For example, in \cite{c23},  a PPG-based CNN-RNN network was designed for 3-stage, 4-stage, and 5-stage classification and achieved accuracy rates of 80.1\%, 68.5\%, and 64.1\%, respectively. In another study \cite{c26}, Light Gradient Boosting Machines and Bi-LSTM networks were used to incorporate multimodal data, such as respiration and wrist accelerometer signals, to enhance classification accuracy. Models achieved up to 79\% accuracy for 4-stage classification.

However, existing machine learning (ML) and DL models fail to achieve robust performance for detailed sleep staging (e.g., 4 or 5 stages) and often require a large volume of datasets or extensive computational resources to train. In addition, The "black-box" nature of most AI models limits their application in clinical settings, as they do not provide explanations of decision-making. Consequently, AI models are difficult to interpret, leaving sleep specialists unable to trust or rely on them in clinical decision-making. Hence, we believe there is a need to develop efficient AI models that utilize wearable data as well as visual interpretability, ensuring clinicians can understand and trust the decisions made by the model.
Thereby, this approach will facilitate the deployment of cost-effective, clinically viable AI-based sleep staging solutions.

In addition to sleep stage classification, sleep disorder classification has been developed in recent years, with studies focusing on various techniques to identify conditions such as insomnia, obstructive sleep apnea, and REM behavior disorder \cite{c24, c25}. Existing research typically employs single-modality approaches, utilizing data such as EEG, ECG, or EMG signals. For example, Zhuang et al. \cite{c24} developed a model with combined EEG, EMG, and ECG signals and reached high sensitivity and specificity for eight types of sleep disorders. Similarly, in another work \cite{c25}, wavelet-based features and ensemble classifiers were deployed to detect six disorders. However, in these studies, sleep disorders are mainly classified as separate entities without consideration for the interrelation between sleep stages and sleep disorders. Despite the complexity of integrating multimodal and multilabel data, simultaneous classification remains an unexplored area. Hence, there is a need to develop an advanced model that can leverage sleep stages and disorders since physiological markers of disorders are often associated with certain sleep stages.

In this paper, we introduce a multitask explainable model based on vision transformer for simultaneous sleep stage and sleep disorder (apnea) classification. The purpose of this approach is to leverage similar underlying characteristics across different sleep disorders and sleep stages. The proposed network incorporates 1D convolution and sequences of patches to capture both local and global dependencies through self-attention mechanisms. We evaluate our model using multimodal data collected from 123 individuals with different sleep disorders. Furthermore, we investigate the contribution of input signals to the final sleep stage and sleep disorder classification by analyzing the self-attention weights from the transformer encoder block. In summary, the paper’s contributions are manifold as follows:
\begin{itemize}
    \item Proposing a 1D-Vision Transformer for simultaneous classification of sleep stage and sleep disorders.
    \item  Evaluating the proposed  model using multimodal- and multilabel data adaptable for home setting environment.
    \item Assessing the proposed model's performance in terms of accuracy, precision, recall, F1-score and Cohen's Kappa.
    \item Investigate the contribution of input signal on the final sleep stage and sleep disorder classification. 
\end{itemize}
The proposed method and dataset are presented in Section 2. The
experimental setup is outlined in Section 3. The method performance is assessed in Section 4. Finally, Section 5 concludes the paper. 
\section{MATERIAL AND METHODS}

\subsection{Data set description}\label{Dataset section}
The dataset used in this study is part of deidentified polysomnography (PSG) recordings from the University of California, Irvine (UCI) Sleep Center \cite{c14}.  The dataset consisted of records pertaining to individual polysomnographic studies on 123 subjects, 48 of whom were diagnosed with obstructive sleep apnea (OSA) and the rest with non-respiratory sleep disorders. Annotations were made every 30 seconds based on analysis of EEG and EMG signals.
The recordings were kept anonymous and were stored in European Data Format (EDF), allowing only the chief diagnosis. All clinical assessments involved at the UCI Sleep Center, conforming to American Academy of Sleep Medicine (AASM) guidelines using the in-lab sleep study with the Natus SleepWorks PSG$^{\circledR}$ acquisition system (Natus Medical Incorporated, Middleton, WI, USA). This study was approved by the Institutional Review Board of UCI (IRB \#267) and the protocols of the same were followed. Sleep staging was delivered by two registered polysomnography technologists, although in quality assurance for interscorer variability, as mandated by the AASM. Each case was reviewed by a board-certified Sleep Physician, with scoring done according to the AASM scoring manual \cite{c13}.

\subsection{Data set preparation}
\begin{figure}[!h]
    \centerline{\includegraphics[height = 8.40cm, width=9.50cm]{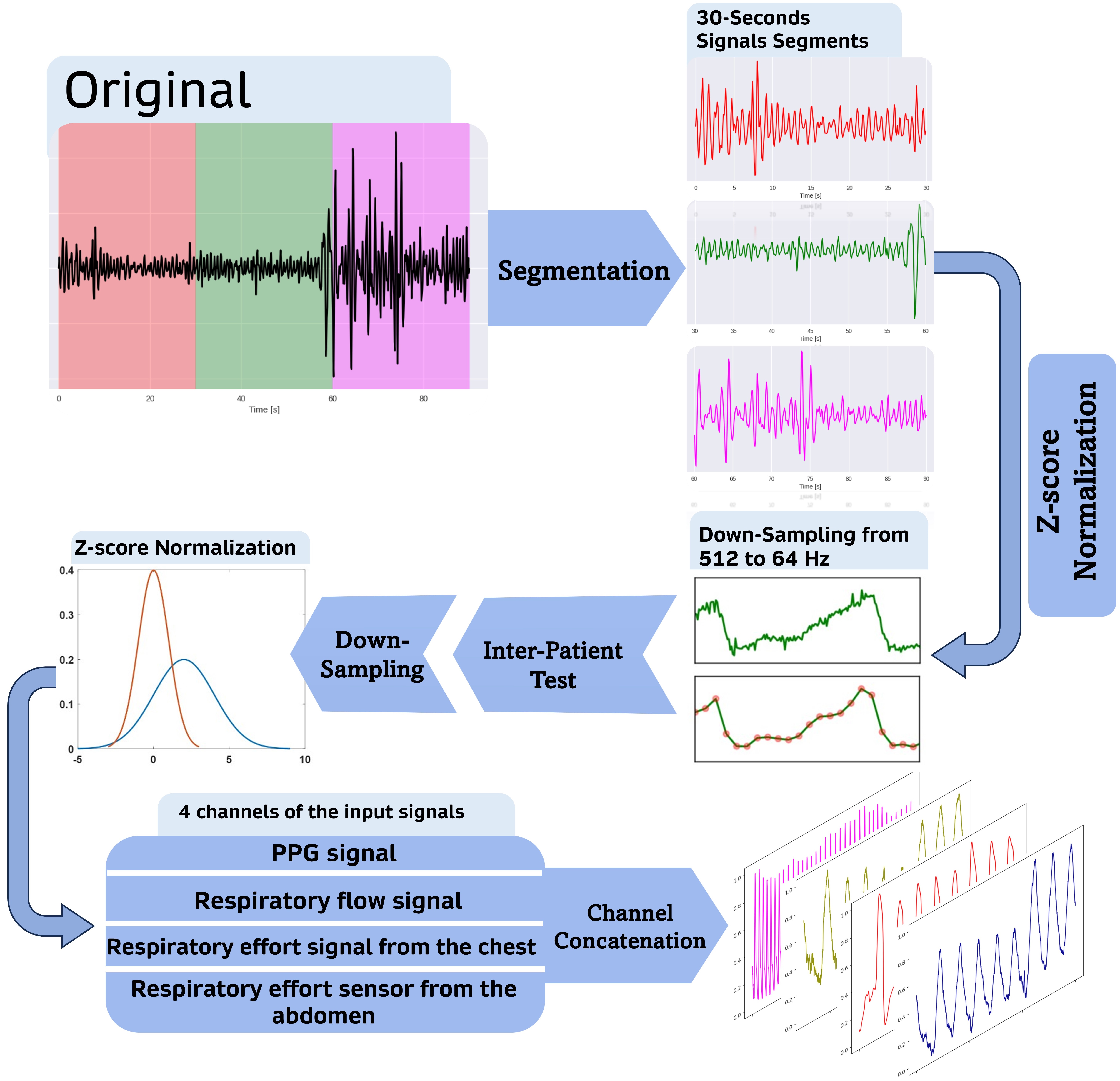}}
    \caption{Data Preparation piepline. The sleep data comprises four channels, representing PPG, respiratory flow signal, chest and abdomen respiratory effort signals. The Signals were segmented into 30-s parts, followed by Z-score normalization on all channels within each subject. Then, an inter-patient test carried out. Next, down-sampling from 512 Hz to 64 Hz was performed on all the signals. Subsequently, the channel concatenation is applied to stack all four channels together. }
    \label{fig1}
\end{figure}

In this section, we outline the data preparation steps (See Fig. \ref{fig1}). Our sleep stages analysis includes four biosignals: the PPG signal from a finger pulse oximeter, respiratory flow signal (RF), and respiratory effort signals placed on the chest and abdomen (RC and RA). In this study, the PPG data is the raw signal that was directly retrieved from the PSG without any additional processing or cleaning.

Data preparation was performed using MATLAB R2022b \cite{c7} (The MathWorks, Natick, Massachusetts, USA). Each data entry included four channels recorded over 30 seconds, with all channels down-sampled from 512 Hz to 64 Hz to make the process more efficient. Thus, we performed a down-sampling method, resulting in each channel containing 1,920 samples per data entry. To achieve this, a simple linear interpolation method \cite{c9} was chosen, which involves connecting data points with straight lines to keep the signal accurate.

To make sure the model's generalizability to new individuals, we used an inter-patient testing approach. We divide the data into the training, validation, and testing groups, each of which consists of different participants. In particular, data from 86 participants were used for training, 18 for validation, and the remaining 18 for testing. 
Next, Z-score normalization was applied to all training data channels. This step was important for helping the model focus on the shape and changes in the signals rather than their actual values. Z-score normalization involved subtracting the mean and dividing by the standard deviation for each channel, which produced data with a mean of zero and a standard deviation of one. This step is especially useful in deep learning because it makes the training process smoother and improves how quickly the model learns.

In the final step, all four channels were stacked to create the full dataset, which had dimensions of 4,994 × 1920 × 4. 4,994 is the total number of 30-second segments, 1920 is the number of samples in each channel, and 4 is the number of channels. The dataset’s sleep stage distribution was as follows: 24.54\% (n = 1,225) for the wake stage, 5.80\% (n = 289) for N1, 40.02\% (n = 2,000) for N2, 15.70\% (n = 788) for N3, and 13.86\% (n = 692) for REM. Furthermore, each 30-second segment has been annotated with the specific type of sleep apnea observed. In total, 2,089 apneas were identified in this study, categorized as follows: Central Apnea (18\%, 372 occurrences), Obstructive Apnea (29\%, 587 occurrences), and Obstructive Hypopnea (53\%, 1,109 occurrences). Having a balanced dataset was crucial for training reliable machine learning models, as it allowed the model to learn from a variety of examples.

\subsection{Multitask learning (MTL)}
In this study, we employ MTL to enhance the performance of classification-based methods by leveraging shared characteristics among classification-related tasks. MTL improves outcomes by capturing correlations between tasks and is often associated with deep learning due to neural networks' ability to learn and generalize shared representations.

A deep learning model is designed to implement MTL for sleep stage and sleep disorder classification. Instead of training separate models for each task, MTL trains a single model to handle multiple interrelated tasks by learning shared representations. This approach enhances generalization and performance, especially when tasks have interdependencies or common features.

During the training phase, MTL uses a joint objective function, optimizing the aggregate loss across tasks. The model processes input data from all tasks simultaneously, adjusting its parameters to extract shared patterns while handling the unique needs of each task.

\subsection{Model Architecture}
\begin{figure}[!h]
    \centerline{\includegraphics[height = 6.60cm, width=9.00cm]{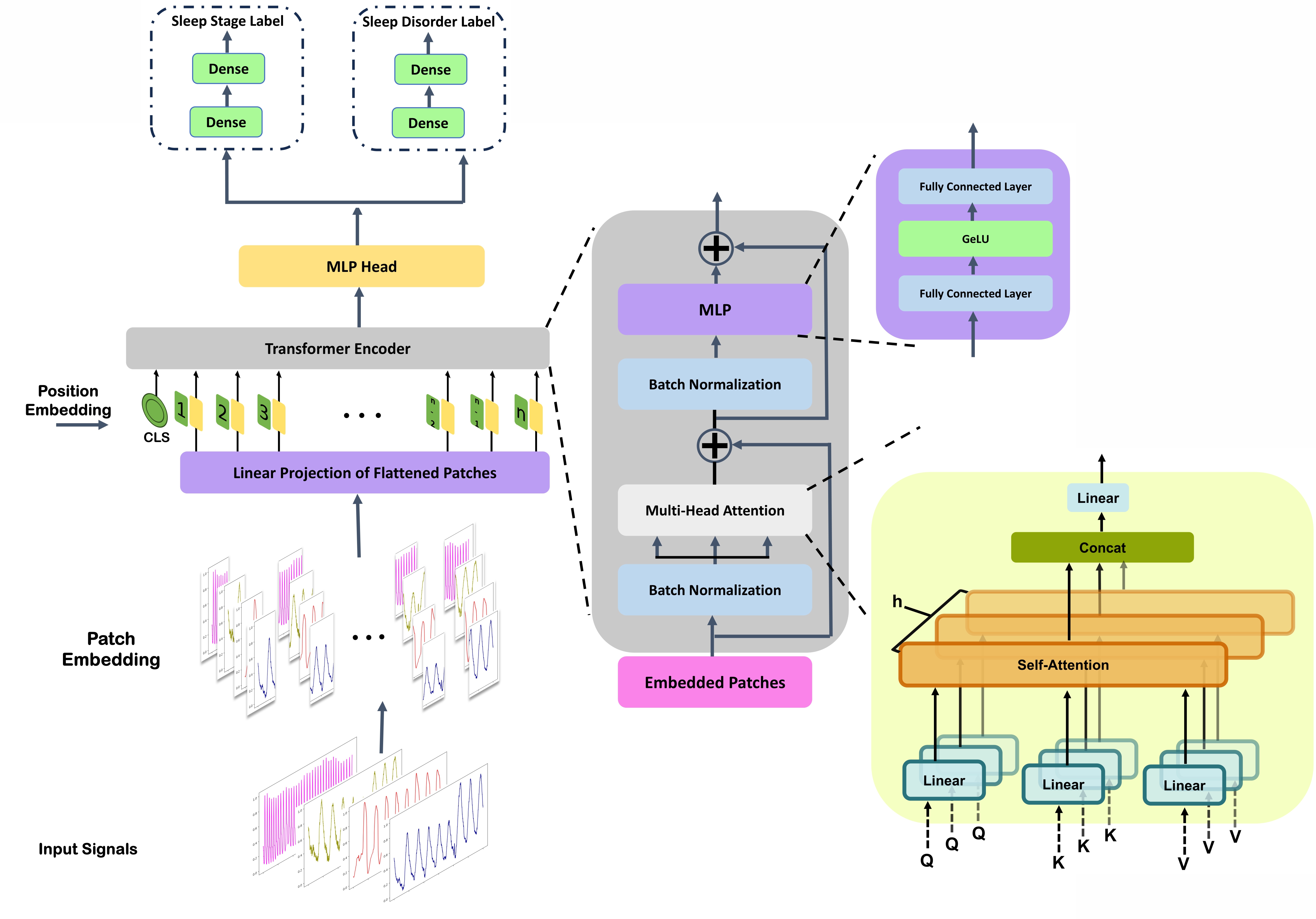}}
    \caption{Illustration of the model architecture. The input signals are divided into patches of a specific size, then linearly embed each patch, incorporate position embeddings, and finally input the sequence of resulting vectors into a standard Transformer encoder. In order to perform classification, an MLP layer is added to the output from the transformer encoder and applies additional transformations to prepare the data for the final classification step. In the final step, two branches of the dense layer are added to perform task-specific (i.e., sleep staging and sleep disorder) classification.} 
    \label{fig2}
\end{figure}

To classify sleep stages and sleep disorders, we developed a one-dimensional Vision Transformer (1D-ViT) model inspired by the original ViT architecture \cite{c10}. Unlike CNNs, the ViT processes input signals as sequences of patches, effectively capturing both local and global dependencies through self-attention mechanisms.
The model architecture, illustrated in Fig. \ref{fig2}, comprises several key components described below.


\subsubsection{Transformer Encoder}
The transformer encoder comprises $L$ layers, each featuring a Multi-Head Self-Attention (MHSA) mechanism and a feedforward neural network (FFN). MHSA enables the model to capture relationships within the input by computing attention scores using Query (Q), Key (K), and Value (V) matrices. Each head focuses on different aspects of the data, and their outputs are concatenated and linearly transformed for the final MHSA block representation. The encoder operations for layer $\ell$ are:
\begin{align}
\mathbf{Z}'_\ell &= \text{LN} \left( \mathbf{Z}_{\ell - 1} + \text{MHSA}(\mathbf{Z}_{\ell - 1}) \right), \\
\mathbf{Z}_\ell &= \text{LN} \left( \mathbf{Z}'_\ell + \text{FFN}(\mathbf{Z}'_\ell) \right),
\end{align}
where $\text{LN}$ denotes layer normalization.
The encoder stabilizes training through layer normalization (LN), residual connections, and dropout, while the GELU activation enhances generalization and performance. After passing through the transformer encoder, the class token $z_0$ is fed into a multi-layer perceptron (MLP) classification head comprising two fully connected layers with 1024 units each. A softmax activation function is applied to MLP layers.



Following the MLP layer, the model includes two output branches, each corresponding to a specific task (sleep stage and sleep disorder classification). The output branches comprised dense layers with ReLU activation functions to capture task-specific features. Finally,  the output layers consist of dense layers with a single neuron and softmax activation functions to predict the probability of the corresponding class for each task (i.e., sleep stage and sleep disorder classification).
\subsection{Training and Optimization}
Training was carried out on 30-second windows of four-channel input data for the classification of five stages of sleep (wake, N1, N2, N3, REM).
A 1D convolutional layer with 768 filters, a kernel size of 20, and a stride of 20 was used to divide input signals into non-overlapping patches. These patch embeddings were then passed through transformer encoder blocks comprising six transformer layers, each with six attention heads and an MLP hidden dimension of 256. To prevent overfitting, dropout was applied at the embedding, attention, and MLP layers. Additionally, a stochastic depth layer with a survival probability of 0.9 was included to regularize the model by probabilistically skipping entire layers during training. For multi-class classification, categorical cross-entropy loss was used:
\begin{equation}
\mathcal{L} = - \frac{1}{N} \sum_{i=1}^{N} \sum_{k=1}^{K} y_{i,k} \log \hat{y}_{i,k},
\end{equation}
where $y_{i,k}$ and $\hat{y}_{i,k}$ are the ground-truth and predicted probabilities for class $k$, respectively.

The AdamW optimizer was employed with a learning rate of $1 \times 10^{-4}$, incorporating weight decay for regularization. Training began with a warm-up learning rate of $1 \times 10^{-4}$ to avoid local minima, which was reduced to $1 \times 10^{-5}$ after 15 epochs for fine-tuning. To address the imbalanced distribution of classes, the focal loss technique is deployed \cite{c12}. This technique reduces the weight of easy samples during training so that the model can focus more on refining its understanding of complex patterns. In this regard, misclassified or challenging samples are given more weight, while correctly classified samples are given less weight. The model was trained for 45 epochs with a batch size of 64, using early stopping based on validation loss to mitigate overfitting.

\subsection{Explainability Integration}
To improve interpretability and understand how the input signal influenced classification decisions, we analyzed the self-attention weights from the final transformer encoder block. Based on the class token and patch token values for each attention head $h$, we extracted the attention scores:
\begin{equation}
\mathbf{a}_h = \mathbf{A}_h[0, 1:N],
\end{equation}
where $\mathbf{a}_h$ represents the contribution of each input data segment to the final prediction. To determine the overall importance of each patch, we averaged the attention scores across all heads:
\begin{equation}
\text{Importance Score}_i = \frac{1}{H} \sum_{h=1}^{H} a_{h,i},
\end{equation}
for $i = 1, \dots, N$. These importance scores were mapped back to the original input signals, allowing us to identify specific segments or events influencing the classification of the model. For example, peaks, troughs, systolic and diastolic regions in PPG signals, or rising times in respiratory signals had the greatest influence on the model's predictions.
Through an examination of how attention was distributed over the input timeline, we determined which physiological components the model emphasized when classifying inputs. This approach highlighted critical patterns in the input signals that contributed to the model's decision-making process.
\section{EXPERIMENTAL SETUP}
The proposed method is evaluated using the dataset described in section \ref{Dataset section}. A total of 123 individuals with different sleep disorders were included in the evaluation. An inter-patient test was carried out by selecting data from individual patients for training, validation and testing sets with a 70\%, 15\%, and 15\% portion. In other words, the training dataset consists of 86 participants (i.e., 3,501 30-s window of 4-channel signals), 18 for validation (i.e., 749 30-s segments of 4-channel signals) and 18 for testing (i.e., 749 30-s segments of 4-channel signals), with no overlap between the three sets. This test ensures that the model has generalization properties and prevents data leakage from the training dataset to the test dataset. In order to enhance computational efficiency, all channels were downsampled from 512 Hz to 64 Hz \cite{c9}. In this method, a line is fitted between each pair of data points, and new data points are generated along the line based on the upsampling or downsampling rate.

\subsection{Evaluation Measures}
In this study, we used five performance scores to describe the model's performance: accuracy, precision, recall, F1-score, and Cohen's Kappa. These metrics are used to measure the difference between the predicted class and the actual class.
In the following each metric is defined:
\begin{equation}
    \textbf{Accuracy} = \frac{\text{Total } \text{number } \text{of } \text{correct } \text{predictions }}{\text{Total } \text{number } \text{of } \text{predictions }}
\end{equation}
\begin{equation}
\textbf{Precision} = \frac{TP }{TP +FP }, \hspace{5mm} \textbf{Recall} = \frac{TP }{TP +FN}
\end{equation}


\begin{equation}
\textbf{$F_1$-score} = \frac{2*\text{Precision }*\text{Recall}}{Precision+Recall}   
\end{equation}

\begin{equation}
 \scalebox{2}{$\kappa$} =  \frac{P_0 - P_e}{1 - P_e}
\end{equation}
where $p_0$ was the observed model accuracy and $p_e$ was the expected
model accuracy. As our classification task is multiclass, we used an average-weighted approach to calculate precision, recall and F1-score, providing a more comprehensive evaluation of the model's performance.
\section{RESULT AND DISCUSSION}
This section describes the performance and effectiveness of the proposed method for sleep stage and sleep disorder classification. In addition, the performance of sleep staging per each sleep disorder is investigated. Following that, the association of each input data channel (PPG, respiratory flow, chest and abdomen respiratory effort ) on classification using ViT attention score is described. Finally, the effect of respiratory disturbances on the classification using the attention scores is investigated.

\subsection{Performance Assessment}\label{Performance Comparison}
This section presents the effectiveness of our proposed MTL model for sleep stage and sleep disorder classification in terms of accuracy, precision, recall, F1-score, and Cohen's Kappa. 
Four-channel input signals are segmented into 30-second intervals. Each segment is annotated with sleep stage labels comprising five classes (Wake, N1, N2, N3, REM) and apnea labels encompassing four distinct types (normal breathing, central apnea, obstructive apnea, and hypopnea). The multitask learning (MTL) framework utilized both label sets simultaneously, aiming to leverage shared features between the tasks to improve overall classification performance.

As shown in Table \ref{tab1}, the model achieved an overall accuracy of 78\% for sleep 5-stage classification, indicating the model correctly classified the majority of samples. Moreover, the model achieved a precision, recall, and F1-score of 0.79 using the four-channel input data, showing balanced and consistent performance across evaluation metrics.

A precision of 0.79 shows that the model makes relatively few incorrect predictions for each sleep stage. In contrast, recall reflects how well a model minimizes false negatives within the total number of instances of a class. A recall of 0.79 demonstrates that the model effectively identifies the most true instances of each class. By combining precision and recall into one metric, the F1 score provides a balanced measure that accounts for both true positives and false negatives. With an F1-score of 0.79, the model achieves a stable trade-off between precision and recall across all sleep stages.

\begin{figure}[h]
\begin{center}
    \centerline{
    \includegraphics[height = 5.10cm, width=7.05cm]{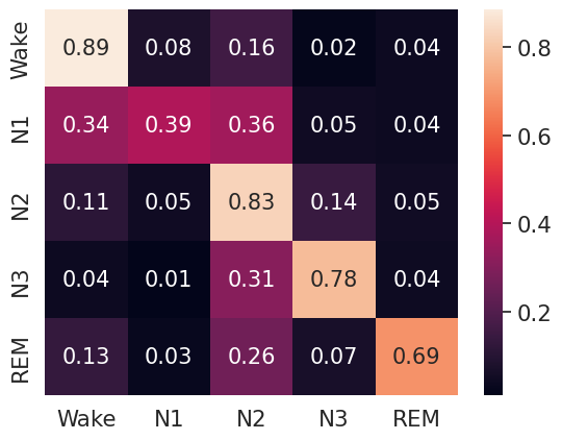}}
    \caption{Confusion Matrix for sleep stage classification. }
    \label{fig3}
\end{center}
\end{figure}

Table \ref{tab2} summarizes the model's performance in sleep disorder classification, which varies across the three primary categories: OSA, hypersomnia, and insomnia. The model achieves an accuracy of 0.94 for hypersomnia, reflecting robust classification and yielding the highest Cohen's Kappa score (0.86). For insomnia, the accuracy is 0.70, indicating moderate performance, while OSA shows the lowest accuracy at 0.56, highlighting the need for improvement, particularly in distinguishing between central and obstructive apneas. The F1-scores per class reveal consistent results for hypersomnia ($\geq$ 0.88 across all classes), whereas OSA classification struggles with lower F1-scores for certain stages, such as N1 (0.28), as shown in Figure \ref{fig3}. As shown in this figure, for REM, the proposed method reached an accuracy of 0.69, with some misclassifications into N2 (0.26) and Wake (0.13). This may be due to the mixed physiological characteristics of REM sleep, which share certain similarities with both Wake and lighter sleep stages.
The results indicate that the model performs well for well-represented and distinct stages like Wake and N2. However, transitional stages like N1 and REM pose significant challenges,  due to overlapping features with adjacent stages.

\begin{table}[!h]
\renewcommand{\arraystretch}{1.10} 
\caption{Performance of sleep stage classification}
\begin{center}
\begin{tabular}{|c|c|c|}

\hline
\multirow{3}{*}{Metrics}    & \multirow{3}{*}{Sleep Stage} & \multirow{2}{*}{Input signal}  \\
                            &                              &                                \\ \cline{3-3} 
                            &                              & \textbf{PPG, RF, RC, RA}  \\ \hline
\textbf{Accuracy}           & \multirow{6}{*}{\textbf{5}}  & 0.78                           \\ \cline{1-1} \cline{3-3} 
\textbf{Precision}          &                              & 0.79                           \\ \cline{1-1} \cline{3-3} 
\textbf{Recall}             &                              & 0.78                           \\ \cline{1-1} \cline{3-3} 
\textbf{F1-score}           &                              & 0.79                           \\ \cline{1-1} \cline{3-3} 
\textbf{F1-score per class} &                              & (0.82, 0.36, 0.79, 0.73, 0.72) \\ \cline{1-1} \cline{3-3} 
\textbf{Kappa}              &                              & 0.66                           \\ \hline

\end{tabular}
\label{tab1}
\end{center}
\end{table}

\begin{table}[!h]
\renewcommand{\arraystretch}{1.1} 
\caption{Performance of sleep stage classification for each sleep disorder type}
\begin{center}
\begin{tabular}{|c|c|ccc|}

\hline
\multirow{5}{*}{Metrics}    & \multirow{5}{*}{Sleep Stage} & \multicolumn{3}{c|}{\multirow{2}{*}{Input signal}}                                                                                                                                                                                       \\
                            &                              & \multicolumn{3}{c|}{}                                                                                                                                                                                                                    \\ \cline{3-5} 
                            &                              & \multicolumn{3}{c|}{\textbf{PPG, RF, RC, RA}}                                                                                                                                                                                       \\ \cline{3-5} 
                            &                              & \multicolumn{3}{c|}{Sleeping Disorder}                                                                                                                                                                                                   \\ \cline{3-5} 
                            &                              & OSA                                                                         & \multicolumn{1}{l}{Hypersomnia}                                              & \multicolumn{1}{l|}{Insomnia}                                               \\ \hline
\textbf{Accuracy}           & \multirow{6}{*}{\textbf{5}}  & 0.56                                                                        & 0.94                                                                         & 0.70                                                                        \\ \cline{1-1} \cline{3-5} 
\textbf{Precision}          &                              & 0.57                                                                        & 0.94                                                                         & 0.70                                                                        \\ \cline{1-1} \cline{3-5} 
\textbf{Recall}             &                              & 0.56                                                                        & 0.95                                                                         & 0.71                                                                        \\ \cline{1-1} \cline{3-5} 
\textbf{F1-score}           &                              & 0.54                                                                        & 0.94                                                                         & 0.70                                                                        \\ \cline{1-1} \cline{3-5} 
\textbf{\begin{tabular}[c]{@{}c@{}}F1-score\\  per class\end{tabular}} &                              & \begin{tabular}[c]{@{}c@{}}(0.72, 0.28,\\ 0.63, 0.61,\\  0.62)\end{tabular} & \begin{tabular}[c]{@{}c@{}}(0.96, 0.42, \\ 0.88, 0.88, \\ 0.88)\end{tabular} & \begin{tabular}[c]{@{}c@{}}(0.79, 0.36, \\ 0.73, 0.78,\\ 0.58)\end{tabular} \\ \cline{1-1} \cline{3-5} 
\textbf{Kappa}              &                              & 0.33                                                                        & 0.86                                                                         & 0.54                                                                        \\ \hline
\end{tabular}
\label{tab2}
\end{center}
\end{table}

\begin{figure}[!htbp]
\begin{center}
    \centerline{
    \includegraphics[width=\linewidth]{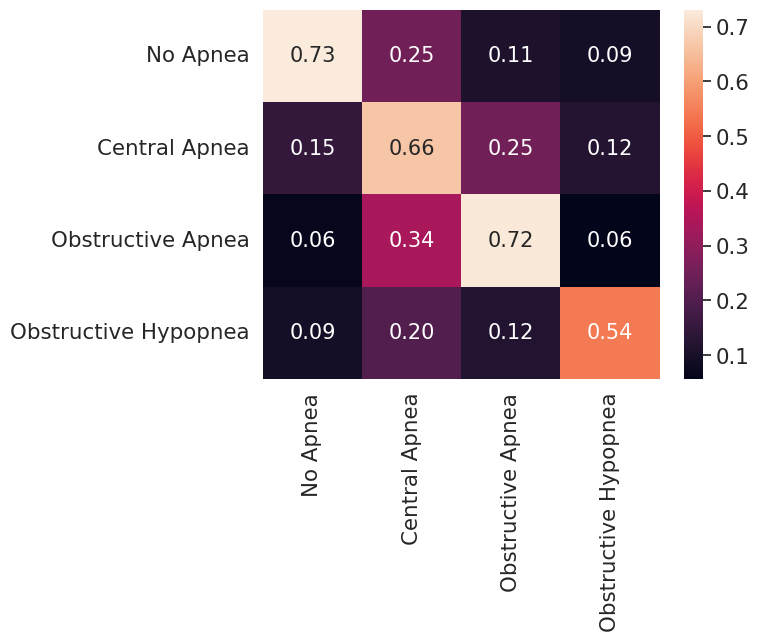}}
    \caption{Confusion Matrix for sleep disorder classification. }
    \label{fig4}
\end{center}
\end{figure}

Table \ref{tab3} shows the summary of the proposed method performance for sleep apnea type classification. As shown in this table, the model achieved an accuracy of 74\%, with a precision of 0.76 and an F1-score of 0.74. Normal breathing was the most accurately classified type (F1-score 0.81), whereas obstructive hypopnea exhibited the lowest F1-score (0.53). Furthermore, the model achieved kappa of 0.58. 
The confusion matrix for sleep apnea classification is illustrated in Figure \ref{fig4}. As shown in this figure, the No Apnea category achieved the highest classification accuracy at 0.73, followed by obstructive apnea and central apnea with accuracy of 0.72 and 0.66, respectively. The lowest accuracy is obtained by obstructive hypopnea with an accuracy of 0.54 with considerable misclassifications into Central Apnea (0.20) and No Apnea (0.09), reflecting the subtle nature of this condition and the difficulty of distinguishing its patterns. Moreover, there is a slight misclassification between obstructive apnea and Central Apnea (0.34), reflecting the physiological similarities between central and obstructive apnea events, such as shared respiratory patterns, likely contributing to this challenge.

\begin{table}[!h]
\renewcommand{\arraystretch}{1.1} 
\caption{Performance of sleep disorder classification}
\begin{center}
\begin{tabular}{|c|c|c|}

\hline
\multirow{3}{*}{Metrics}    & \multirow{3}{*}{Sleep Apnea type} & \multirow{2}{*}{Input signal}  \\
                            &                              &                                \\ \cline{3-3} 
                            &                              & \textbf{PPG, RF, RC, RA}  \\ \hline
\textbf{Accuracy}           & \multirow{6}{*}{\textbf{4}}  & 0.74                           \\ \cline{1-1} \cline{3-3} 
\textbf{Precision}          &                              & 0.76                           \\ \cline{1-1} \cline{3-3} 
\textbf{Recall}             &                              & 0.73                           \\ \cline{1-1} \cline{3-3} 
\textbf{F1-score}           &                              & 0.74                           \\ \cline{1-1} \cline{3-3} 
\textbf{F1-score per class} &                              & (0.81, 0.67, 0.68, 0.53) \\ \cline{1-1} \cline{3-3} 
\textbf{Kappa}              &                              & 0.58                           \\ \hline

\end{tabular}
\label{tab3}
\end{center}
\end{table}

In summary, the proposed MTL framework demonstrated strong performance in both sleep stage and sleep disorder classification. It effectively leveraged their correlations to enhance overall classification accuracy. The model combines multimodal data to achieve strong accuracy in sleep staging while also identifying key patterns associated with sleep disorders. Compared to traditional single-task models, MTL-based approaches offer a more comprehensive and efficient method for improving automated sleep assessment.
\subsection{Model Explainibility}
This section explores the attention scores generated by the ViT model in sleep classification. The attention distributions are evaluated across different components of the input signals. To achieve this, the trained model is provided with input signals (PPG, RF, RC and RA) alongside their corresponding sleep stage and sleep apnea labels. The signals are processed through all layers of the encoder except for the final block, which extracts the attention scores from each self-attention head. These scores are then visualized over the input signals, highlighting the significance of specific signal segments as determined by the ViT model.

\begin{figure}[ht!]
    \begin{subfigure}[b]{0.5\textwidth}
        \centering
        \includegraphics[height=1.70cm, width=\textwidth]{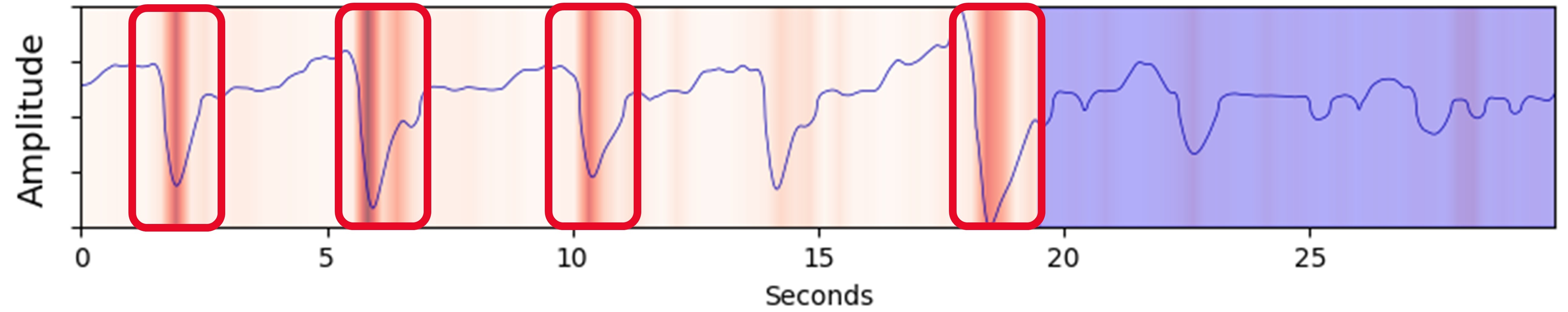}
        \vspace{-1.2em}
        \caption{}
        \label{fig:OH}
    \end{subfigure}
    \hfill
    \begin{subfigure}[b]{0.5\textwidth}
        \centering
        \includegraphics[height=1.70cm, width=\textwidth]{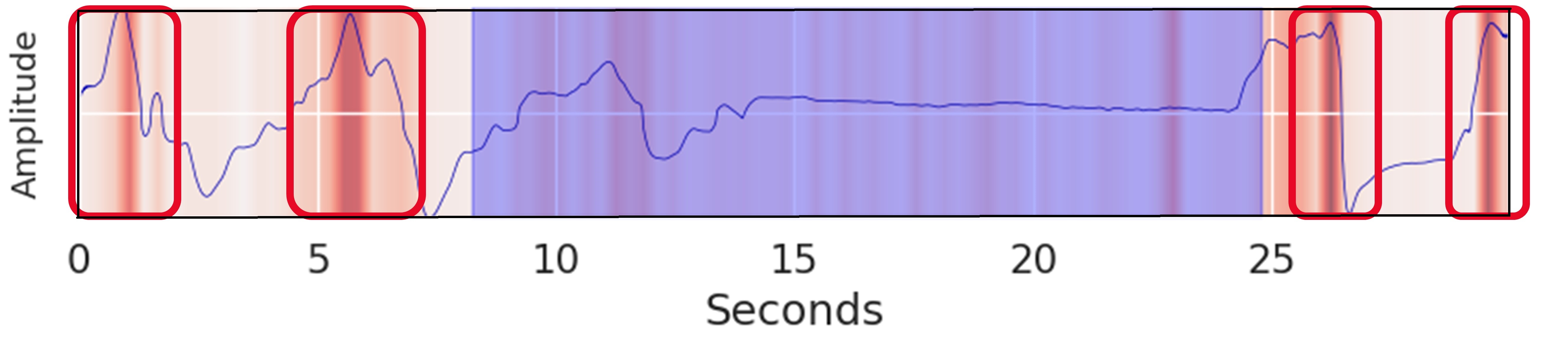}
        \vspace{-1.2em}
        \caption{}
        \label{fig:OH1}
    \end{subfigure}
    \hfill
    \begin{subfigure}[b]{0.5\textwidth}
        \centering
        \includegraphics[height=1.70cm, width=\textwidth]{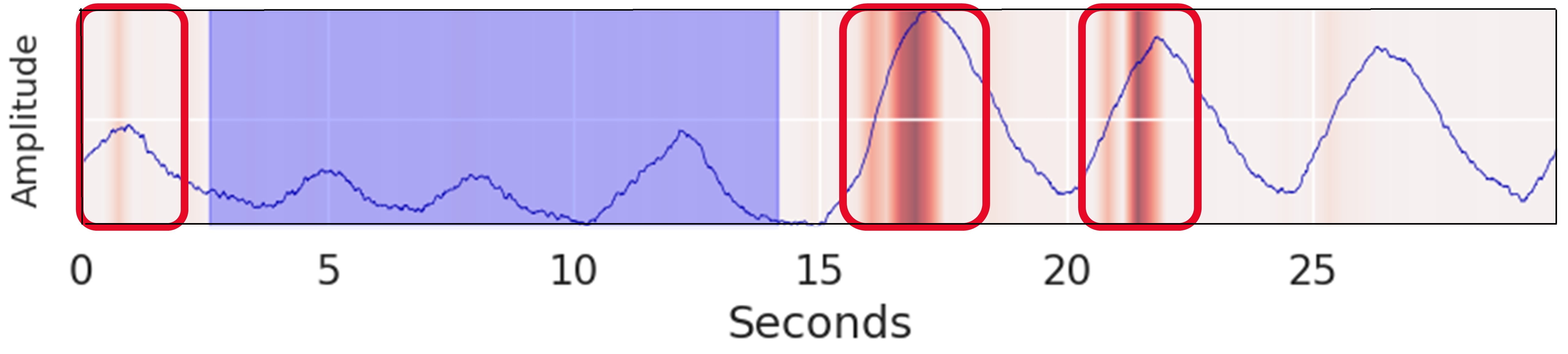}
        \vspace{-1.2em}
        \caption{}
        \label{fig:OH2}
    \end{subfigure}
    \hfill
    \begin{subfigure}[b]{0.5\textwidth}
        \centering
        \includegraphics[height=1.70cm, width=\textwidth]{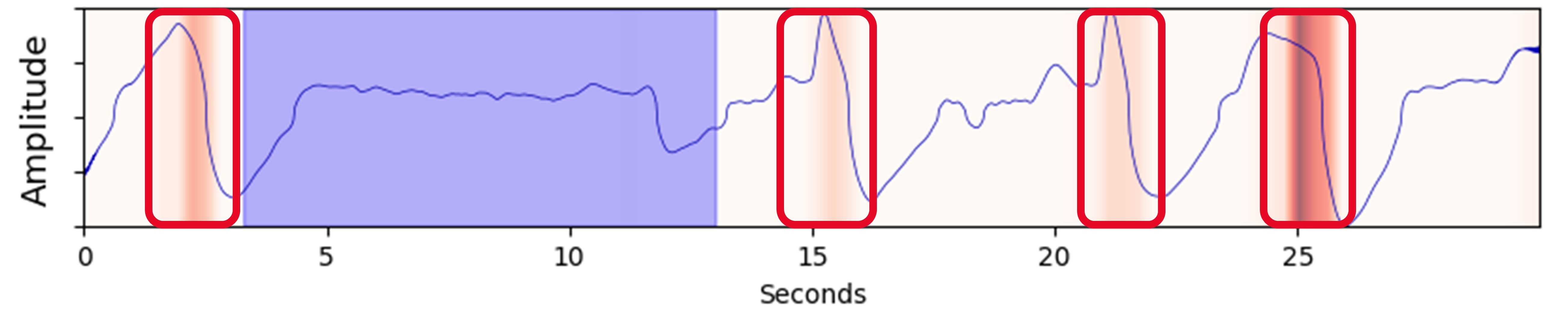}
        \vspace{-1.2em}
        \caption{}
        \label{fig:CA}
    \end{subfigure}
    \hfill
    \begin{subfigure}[b]{0.5\textwidth}
        \centering
        \includegraphics[height=1.70cm, width=\textwidth]{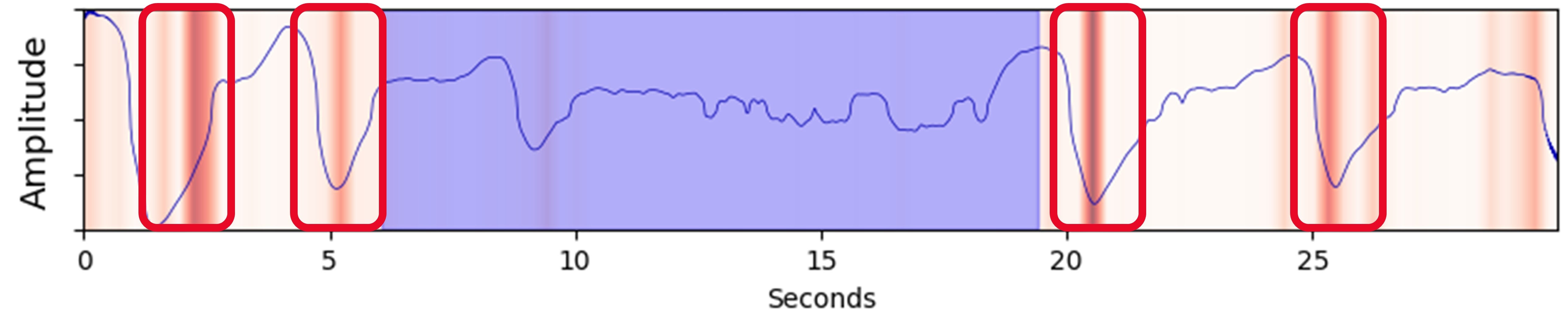}
        \vspace{-1.2em}
        \caption{}
        \label{fig:OSA}
    \end{subfigure}

    \caption{Attention Maps plotted on top of respiratory signal for different sleep disorder occurrence (a) Obstructive Hypopnea, (b) Obstructive Hypopnea, (c) Obstructive Hypopnea, (d) Central Apnea and (e) Obstructive Sleep Apnea. The high score attentions are colored in darker red, and the actual respiration signal is illustrated in solid blue line. The sleep disturbance is shadowed in blue.}
    \label{fig:att}
\end{figure}

For example, in Fig. \ref{fig:att}, the respiration signals alongside the distribution of the attention scores are plotted. Each figure highlights an apnea event, represented by a shadowed blue area. As shown in these figures, the attention maps demonstrate that the model focuses on specific features of the respiratory signals, such as troughs, peaks, and inhalation/exhalation phases, which play a crucial role in differentiating between sleep stages. For instance, in Fig \ref{fig:OH} and \ref{fig:OSA}, the shadowed blue area illustrates an obstructive apnea and obstructive sleep apnea event occurring toward the end of the 30-second respiratory signal, respectively. In these figures, the attention maps reveal that the model prioritizes specific respiratory signal features such as troughs, which are critical for distinguishing sleep stages. Respiratory troughs, which occur during the shift from exhaling to inhaling, show changes in breathing patterns that vary depending on the sleep stage. In NREM sleep, especially in the deeper stages like NREM Stage 3, these troughs are deeper and more consistent, reflecting steady and regular breathing. On the other hand, during REM sleep, the troughs are shallower and more uneven due to greater fluctuations in the activity of the autonomic nervous system \cite{c15, c16}. Similarly, as depicted in Fig. \ref{fig:OH1}, \ref{fig:OH2}, and \ref{fig:CA}, respiration peaks, representing the highest point of inhalation, are more pronounced and stable during NREM sleep. In contrast, REM sleep is characterized by peaks that are less predictable and more variable \cite{c15}. 

By focusing on these physiologically relevant events, the model effectively identifies the underlying patterns associated with each sleep stage. This approach aligns with the established respiratory characteristics \cite{c17} in sleep staging and demonstrates the model's ability to extract meaningful features directly from the raw signal without requiring manual feature engineering. Furthermore, the attention map shows that the model assigns comparatively less attention to signal regions affected by apneas. Apneic events, typically associated with sleep-disordered breathing, are known to disrupt normal respiratory patterns \cite{c17}.  By placing less emphasis on these regions for sleep staging, the model reduces the risk of relying too much on irregular or pathological patterns that may not consistently indicate specific sleep stages across different individuals. However, despite the reduced attention on these segments, the model still processes them sufficiently to recognize and classify apneic events. This balanced and adaptive approach enables the model to perform multitasking effectively, leveraging physiological insights to achieve high interpretability and robust performance.






\section{CONCLUSIONS}
In this study, we proposed an MTL framework utilizing a 1-D Vision Transformer for simultaneous sleep stage and sleep disorder classification using multi-modal data. The proposed MTL method was implemented to leverage the potential interdependencies and shared characteristics between sleep stages and sleep disorders since physiological indicators of disorders are typically associated with certain stages of sleep. The method was evaluated using a dataset collected from 123 subjects who were diagnosed with OSA and non-respiratory sleep disorders.  Experimental results demonstrated that MTL models achieved an accuracy of 78\% for five-stage sleep classification and 74\% for sleep apnea classification. Furthermore, by analyzing attention weights, we demonstrated that different events in the signal segments, such as respiratory troughs and peaks, have a higher contribution in the final classification process.
These results highlight the potential of our approach to balance performance in sleep staging while simultaneously identifying sleep disorders, thereby improving noninvasive, home-based sleep monitoring.

\addtolength{\textheight}{-12cm}   



\section*{Acknowledgement}
This work was partially supported by the Finnish Foundation for Technology Promotion and the Nokia Foundation.
\section*{Disclosing Financial Interests}
The authors declare the following financial interests/personal relationships which may be considered as potential competing interests: Michelle Khine reports a relationship with Vena-Vitals that includes: equity or stocks.


\end{document}